# Directional strong coupling at the nanoscale between hyperbolic polaritons and organic molecules

A. I. F. Tresguerres-Mata[1,2†], O. G. Matveeva[3†], C. Lanza[1,2†], J. Álvarez-Cuervo[1,2], K. V. Voronin[3,4], F. Calavalle[5], G. Avedissian[5], P. Díaz-Núñez[6,7], G. Álvarez-Pérez[1,2,8], A. Tarazaga Martín-Luengo[1,2], J. Taboada-Gutiérrez[9], J. Duan[1,2,10,11], J. Martín-Sánchez[1,2], A. Bylinkin[5], R. Hillenbrand[5,12], A. Mishchenko[6,7], Luis E. Hueso[5,12], V. S. Volkov[13], A. Y. Nikitin[3,12*], and P. Alonso-González[1,2*]

[1]*Department of Physics, University of Oviedo, Oviedo 33006, Spain.*
[2]*Center of Research on Nanomaterials and Nanotechnology, CINN (CSIC-Universidad de Oviedo), El Entrego 33940, Spain.*
[3]*Donostia International Physics Center (DIPC), Donostia/San Sebastián 20018, Spain.*
[4]*Universidad del País Vasco/Euskal Herriko Unibertsitatea, Donostia/San Sebastián 20080, Basque Country, Spain.*
[5]*CIC nanoGUNE BRTA, 20018 Donostia/San Sebastián, Spain.*
[6]*Department of Physics and Astronomy, University of Manchester, Manchester, UK.*
[7]*National Graphene Institute, University of Manchester, Manchester, UK.*
[8]*Present address: Center for Biomolecular Nanotechnologies, Istituto Italiano di Tecnologia, via Barsanti 14, Arnesano, 73010, Italy.*
[9]*Department of Quantum Matter Physics, Université de Genève, 24 Quai Ernest Ansermet, CH-1211, Geneva, Switzerland.*
[10]*Present address: Centre for Quantum Physics, Key Laboratory of Advanced Optoelectronic Quantum Architecture and Measurement (MOE), School of Physics, Beijing Institute of Technology, Beijing 100081, China.*
[11]*Present address: Beijing Key Laboratory of Nanophotonics & Ultrafine Optoelectronic Systems, Beijing Institute of Technology, Beijing 100081, China.*
[12]*IKERBASQUE, Basque Foundation for Science, Bilbao 48013, Spain.*
[13]*XPANCEO, Bayan Business Center, DIP, 607-0406, Dubai, UAE.*

*pabloalonso@uniovi.es  alexey@dipc.org
[†] *These authors contributed equally to this work.*

**Strong coupling (SC) is a fundamental concept in physics that describes extreme interactions between light and matter. Recent experiments have demonstrated SC at the nanometer scale, where strongly confined polaritons, rather than photons, couple to quantum emitters or molecular vibrations. Coupling with the latter is generally referred to as vibrational SC (VSC) and is of significant fundamental and technological interest, as it can be an effective tool for modifying molecular properties. However, the implementation of VSC, especially at the nanoscale, depends on the development of tuning mechanisms that allow control over the coupling strength and, eventually, its directionality, opening the door for the selective coupling of specific molecular vibrations. Here we report the observation of directional VSC. Specifically, we show nanoscale images of propagating hyperbolic phonon polaritons (PhPs) coupled to pentacene molecules revealing that the fingerprint of VSC for propagating polaritons –a marked anti-crossing in their dispersion at the vibrational resonance- can be modulated as a function of the direction of propagation. In addition, we show that VSC can exhibit an optimal condition for thin molecular layers, characterized by a maximum coupling strength along one single direction. This phenomenon is understood by analysing the overlap of the**

**polariton field with molecular layers of varying thicknesses. Apart from their fundamental importance, our findings promise novel applications for directional sensing or local directional control of chemical properties at the nanoscale.**

In condensed matter physics, strong coupling (SC) describes an intense interaction of light and matter such that the energy exchange between them exceeds the decoherence rates, leading to the formation of hybridized states. This concept has been studied extensively and has been successfully applied, for example, to tune the emission of quantum emitters coupled to microcavities[1,2]. Importantly, the existence of SC between light and molecular vibrations (VSC) has also been demonstrated, which has attracted much attention because it promises the ability to manipulate fundamental molecular properties such as chemical reactivity, selectivity, and photostability[3-7]. Recent experiments have extended these studies to the nanoscale by exciting strongly confined surface polaritons[8-10] instead of photons[11-14]. Among them, those using propagating surface polaritons are of particular interest[11], since they allow combining low optical losses and mid-infrared frequencies, where a great variety of molecular vibrations reside. However, to fully exploit the potential of VSC, it is still necessary to implement tuning mechanisms that allow precise control of the coupling strength as well as the direction in which it occurs. Such directionality would potentially allow the selective coupling of specific molecular vibrations, thus facilitating the understanding and prediction of the behaviour of molecules, reactions, and materials[3-7]. Despite this great potential, the concept of directionality remains largely unexplored in the framework of the SC regime.

In this work, we report the observation of directional VSC. Specifically, we demonstrate that the SC between propagating PhPs and nearby molecules can be modulated/controlled by the in-plane direction along which the former propagate at the nanoscale. Furthermore, we find that VSC can be designed to exhibit a maximum value along one single direction.

The basis of our experiment is shown in Fig. 1a, together with a schematic of the main result obtained (top panel). We consider an α-$MoO_3$ flake, which supports in-plane hyperbolic PhPs at mid-IR frequencies (between 821 $cm^{-1}$ and 963 $cm^{-1}$)[15-18], placed on top of a layer of isotropic pentacene molecules[19] (see Methods), which exhibit strong vibrational absorption at mid-IR frequencies (γ(C-H) mode at 904 $cm^{-1}$)[20]. In particular, we study the propagation of PhPs in α-$MoO_3$ along different in-plane directions, and thus with different dispersion, in the presence of the molecules. To do this, we carry out polariton interferometry employing scattering-type Scanning Near-field Optical Microscopy (s-SNOM). This technique has been recently established to visualize SC phenomena in real space[11]. As depicted in the figure, the s-SNOM employs a metallic-coated AFM tip, which scans the surface of the sample while being illuminated by an incident field, $E_{inc}$. The tip launches PhPs, which propagate towards the edge of the flake and get reflected back to the tip. The resulting PhP field is further backscattered out of plane ($E_{sca}$) towards the detector (technique known as polariton interferometry[21-23]). Importantly, the α-$MoO_3$ crystal flake is selected to exhibit a naturally smooth and continuous curved edge shape (see optical image in Fig. 1b). This geometry allows the polariton interferometry technique to analyse the polariton dispersion as a function of in-plane direction since there are reflecting edges for a wide set of in-plane angles. In the presence of molecules, these measurements will therefore allow us to study the directionality (if any) of the polariton-molecular vibration coupling.

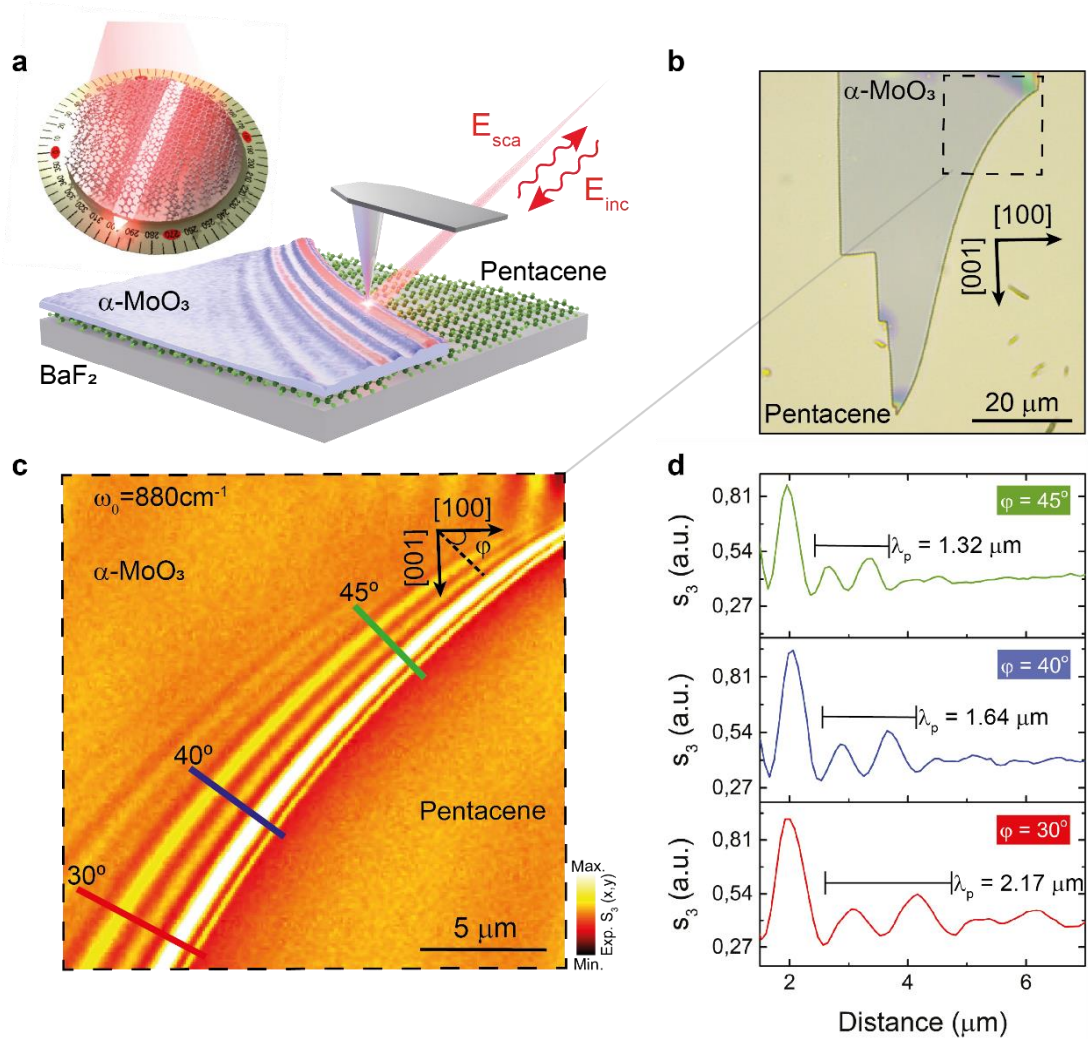

**Fig. 1. Observation of directional VSC by imaging the in-plane anisotropic propagation of PhPs in an α-MoO₃ thin layer placed on pentacene molecules. a**, Illustration of directional VSC. Our experiment consists of s-SNOM imaging of PhPs propagating anisotropically in the plane of an α-MoO₃ thin layer placed on pentacene molecules. **b**, Optical image of the α-MoO₃ thin layer (200 nm in thickness) with a curved edge placed on top of pentacene molecules (36 nm in thickness). **c**, Near-field amplitude s-SNOM image of the α-MoO₃/pentacene area indicated in (b) by a dashed square. The image is taken at an illumination frequency of 880 cm⁻¹. **d**, Line profiles of the near-field amplitude image in (c) showing the propagation of PhPs along three different in-plane directions (in-plane angle φ): 45° (green), 40° (blue), and 30° (red). The polaritonic wavelength decreases with increasing in-plane angle, demonstrating the strong in-plane anisotropic propagation of PhPs in α-MoO₃.

First, we perform polariton interferometry of our α-MoO₃/pentacene sample at a frequency of 880 cm⁻¹ (i.e. $\lambda_{inc}$=11.36 μm), which is spectrally far from the molecular resonant absorption at 904 cm⁻¹. The resulting near-field image (Fig. 1c) shows bright fringes parallel to the curved edge of the α-MoO₃ flake exhibiting different periodicities for different in-plane directions. This observation clearly indicates an in-plane anisotropic propagation of PhPs in the sample, which can be further corroborated by extracting linear profiles along three different in-plane directions with respect to the [100] crystalline direction, for example at φ = 45°, 40°, and 30° (green, blue, and red curves in Fig. 1d).

We note that the wavelength of the PhPs strongly decreases (from 2.2 μm to 1.3 μm) with increasing φ, implying higher wavelength confinement (from $\lambda_{inc}/5$ to $\lambda_{inc}/9$).

Next, we study the effect of the molecular resonant absorption on the anisotropic propagation of PhPs in α-MoO$_3$. To do this, we take analogous near-field images at several different frequencies that now also cover the molecular absorption and extract the polaritonic dispersion by measuring the PhPs wavelength as a function of frequency (see Supplementary Note I) for different in-plane directions (Fig. 2).

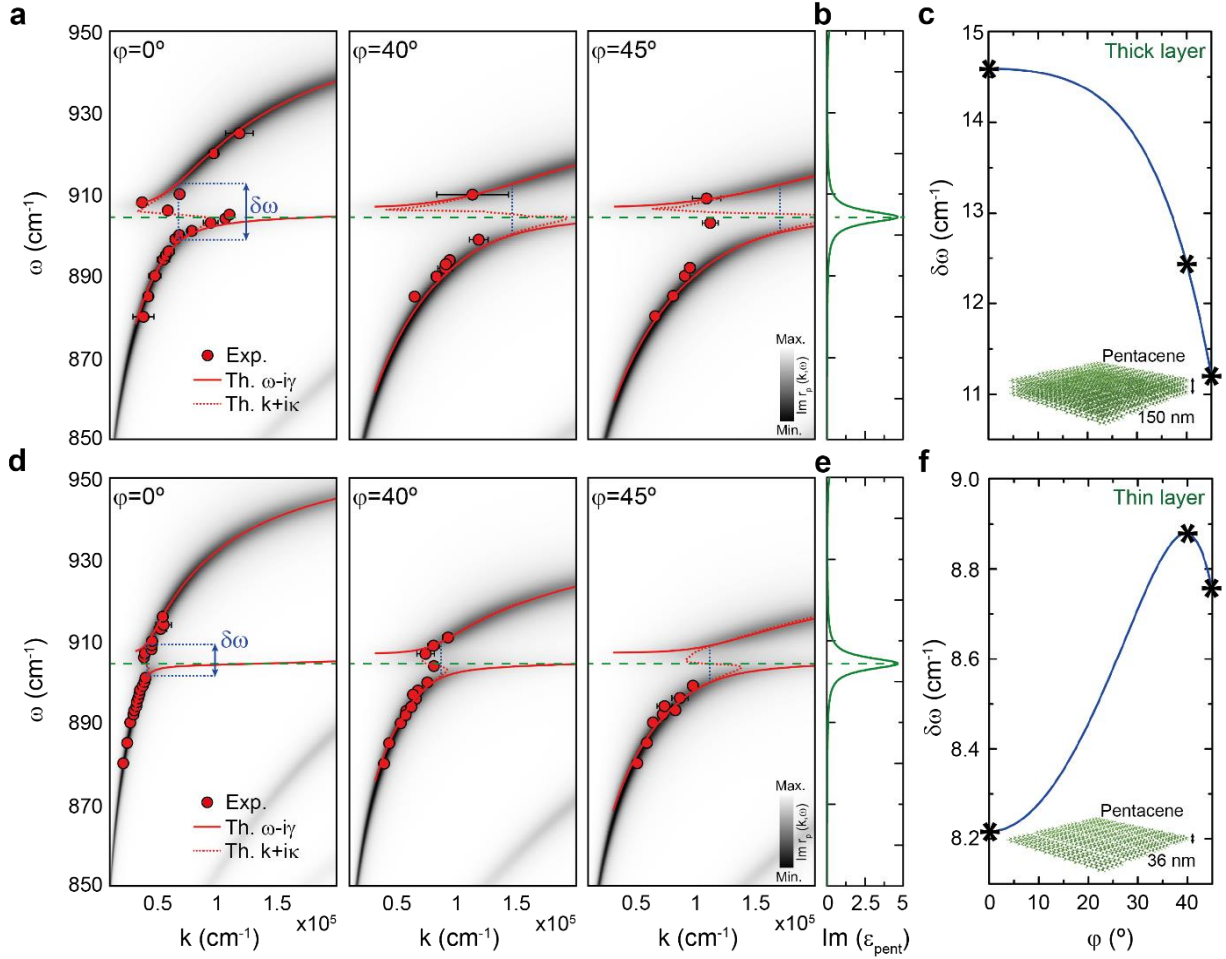

**Fig. 2. Directionally-modulated VSC between propagating PhPs in α-MoO$_3$ and pentacene molecules. a**, Experimental (red dots) and theoretical dispersions (color plot from TM and red solid/red dashed lines from analytical calculations assuming complex frequencies/complex momenta, ω-iγ and k+iκ, respectively) of PhPs in α-MoO$_3$ (135 nm in thickness) on a 150 nm-thick pentacene layer for 0° (left), 40° (middle), and 45° (right) in-plane angles. δω (blue dashed lines) indicates the separation between branches (mode splitting). **b**, Imaginary part of the permittivity of the pentacene molecule (Im (ε$_{pent}$)), showing a peak at the absorption frequency of the pentacene molecule at 904 cm$^{-1}$. Its maximum indicates the frequency at which the dispersions exhibit mode splitting (green dashed line). **c**, Analytical calculation (assuming complex frequencies) of δω (blue line) for the case shown in (a), together with the experimentally extracted values (black symbols). A clear decrease of δω is observed with increasing in-plane angle. However, the SC condition defined as $C \stackrel{\text{def}}{=} \frac{\delta\omega^2}{\frac{\Gamma_{PhP}^2}{2}+\frac{\Gamma_{mol}^2}{2}} > 1$ is satisfied in all cases. **d**, Same as (a) for an α-MoO$_3$ flake (200 nm in thickness) placed on a 36 nm-thick pentacene layer. **e**, Same as (b). **f**, Analytical calculation (assuming complex frequencies) of δω (blue line) for the case shown

in (b), together with the experimentally extracted values (black symbols). In this case, δω shows a resonance behaviour with a maximum at φ= 40º. As in (c), the SC condition is fulfilled in all cases.

Fig. 2a shows the resulting dispersions (red dots) for a sample composed of a 135 nm thick α-MoO$_3$ flake placed on a relatively thick pentacene layer (150 nm). Three different in-plane angles are examined (φ=0º, 40º, and 45º, left, middle and right panels, respectively). As a general trend, we observe that the polariton wavelength decreases (i.e., the wave vector k or polariton momentum increases) with increasing frequencies for all three angles. However, at the absorption frequency of the pentacene molecule (904 cm$^{-1}$ shown in Fig. 2b as the Im($\varepsilon_{pent}$)), we observe a very different behaviour in the PhP dispersion, with the polariton momenta k decreasing with frequency. This effect introduces a back-bending in the dispersion, unlike what is typically observed in an α-MoO$_3$[16] flake without pentacene molecules (see Supplementary Note II). In addition, in the frequency range covered by such back-bending (~ 903-906 cm$^{-1}$) we observe a strongly reduced propagation length (see Supplementary Note III). Remarkably, beyond the absorption of the molecule, the dispersion recovers the initial trend with the polariton momenta increasing with frequency. As a result, the experimental dispersion can be described by a continuous dispersion displaying a back-bending region at the vibrational resonance of the pentacene molecule. This observation is a clear signature of light-matter coupling, as previously reported[11,24], occurring for all three in-plane angles studied. Interestingly, there are, however, clear differences in the back-bending region for the different angles. Most notably, the larger the angle, the larger the momenta at which the back-bending occurs.

To better understand and quantify the dependence of the coupling on the in-plane angle, we performed a detailed theoretical analysis. Specifically, we first calculated numerically the polariton dispersion using the transfer matrix (TM) method[25]. The resulting plots (grey color plots in Fig. 2a) match well with the experimental data. In addition, we calculated the quasi-normal modes of the α-MoO$_3$/pentacene system using an analytical approach[26]. In these calculations we assume either complex frequencies, ω-iγ, i.e., considering a temporally-decaying wave, or complex momenta, k+iκ, i.e., considering a spatially-decaying wave (see Supplementary Note IV). Both explanations are describing the same physical system but focus on different aspects of the behaviour of the modes within it. Under the latter assumption (spatially-decaying wave), the calculated curves show a continuous dispersion with a clear back-bending in the pentacene absorption resonance (dashed red curves in Fig. 2a), aligning very well with the experimental data for the three angles studied (left, middle and right panels for φ=0º, 40º, and 45º, respectively). The back-bending suggests that the modes experience strong interactions that modify their dispersion (anomalous dispersion with strongly reduced propagation lengths) resulting in a continuous curve rather than distinct branches.

Once corroborated that our model reproduces well the experimental results, we can try now to analyse if SC is reached in our system and, more importantly if it shows any dependence with the in-plane angle. To do this, we plot in Fig 2a the dispersions calculated with our model considering complex frequencies (solid red curves), which, unlike the previously used representation, allows us to accurately extract the δω at the value of k where the distance is minimal. Temporally decaying modes are characterized by the decrease in amplitude over time. In the SC regime, these modes exhibit anti-crossing and mode splitting, resulting in two separate branches from which we can obtain δω. Traditionally, SC is characterized by the relation between the splitting between dispersion branches (δω) and the resonance linewidths (Γ), which can be quantify according to the expression $C \stackrel{\text{def}}{=} \frac{\delta\omega^2}{\frac{\Gamma_{PhP}^2}{2}+\frac{\Gamma_{mol}^2}{2}} > 1$. The resulting values of δω in our experiment (Fig. 2c) indicate that the SC condition[27] is satisfied for all in-plane angles (see Supplementary Note II). In fact, the limiting case indicating a system that is not strongly coupled, would correspond to a value of δω = 3.7 cm$^{-1}$ for the 0° case and δω = 4.1 cm$^{-1}$ for the 40° case. We note that the uncoupled PhPs linewidth is angle dependent such that $\Gamma_{PhP}$ = 3.8 cm$^{-1}$ and 4.6 cm$^{-1}$ for φ=0° and 40°, respectively. On the other hand, $\Gamma_{mol}$ is 3.5cm$^{-1}$ in all cases (see Supplementary Note V). More importantly, the evolution of δω with the in-plane angle exhibits a clear variation, reducing its value from approximately 14.5 cm$^{-1}$ at φ = 0° to 11 cm$^{-1}$ at φ = 45°, revealing that the SC condition is, in turn, directionally modulated. This observation thus establishes the concept of directionality in the framework of VSC phenomena.

To better understand the physics behind our observations, we studied the dependence of the coupling on the thickness of the molecular layer. To this end, we placed a 200 nm thick α-MoO$_3$ flake on a 36 nm-thick pentacene layer and reconstructed the polaritonic dispersion from the near-field images as a function of the in-plane angle (red dots in the left, middle, and right panels of Fig. 2d for φ=0°, 40° and 45°, respectively). As in the previous case, we observe a clear back-bending effect in the dispersion curve for all in-plane angles at the absorption frequency (904 cm$^{-1}$) of the pentacene molecule (Fig. 2e). This behaviour of the experimental polaritonic dispersions closely follows the trend of those numerically calculated by TM formalism (grey color plots) or analytically in complex momentum (solid red curves) and complex frequency (dashed red curves). For the latter, we observe an excellent agreement between experiment and theory through back-bending of the dispersion, as previously observed for the thicker pentacene layer. We also observe, as before, that the dispersions shift towards larger momenta as the in-plane angle increases. However, when comparing the dispersions for different molecular layer thicknesses (Fig. 2a and 2d), we observe that the separation between the upper and lower branch, δω, indicative of the coupling strength, is smaller for the thinner layer. To quantify this observation, we extract δω from the complex frequency plots (Fig. 2f), obtaining values ranging between 8.2 and 8.9 cm$^{-1}$, which are effectively smaller than in the previous case. However, although smaller, these values also reveal SC for all in-plane angles, since the condition $C > 1$ is again satisfied, indicating that we are always above the limiting case. Importantly, unlike the case of the thicker pentacene layer, we now observe a strongly non-monotonous trend for δω, increasing until reaching a maximum at around φ=40° and then decreasing. This observation, apart from corroborating a directional SC behaviour, unveils the existence of an optimal condition (optimal in-plane direction) for realizing VSC experiments.

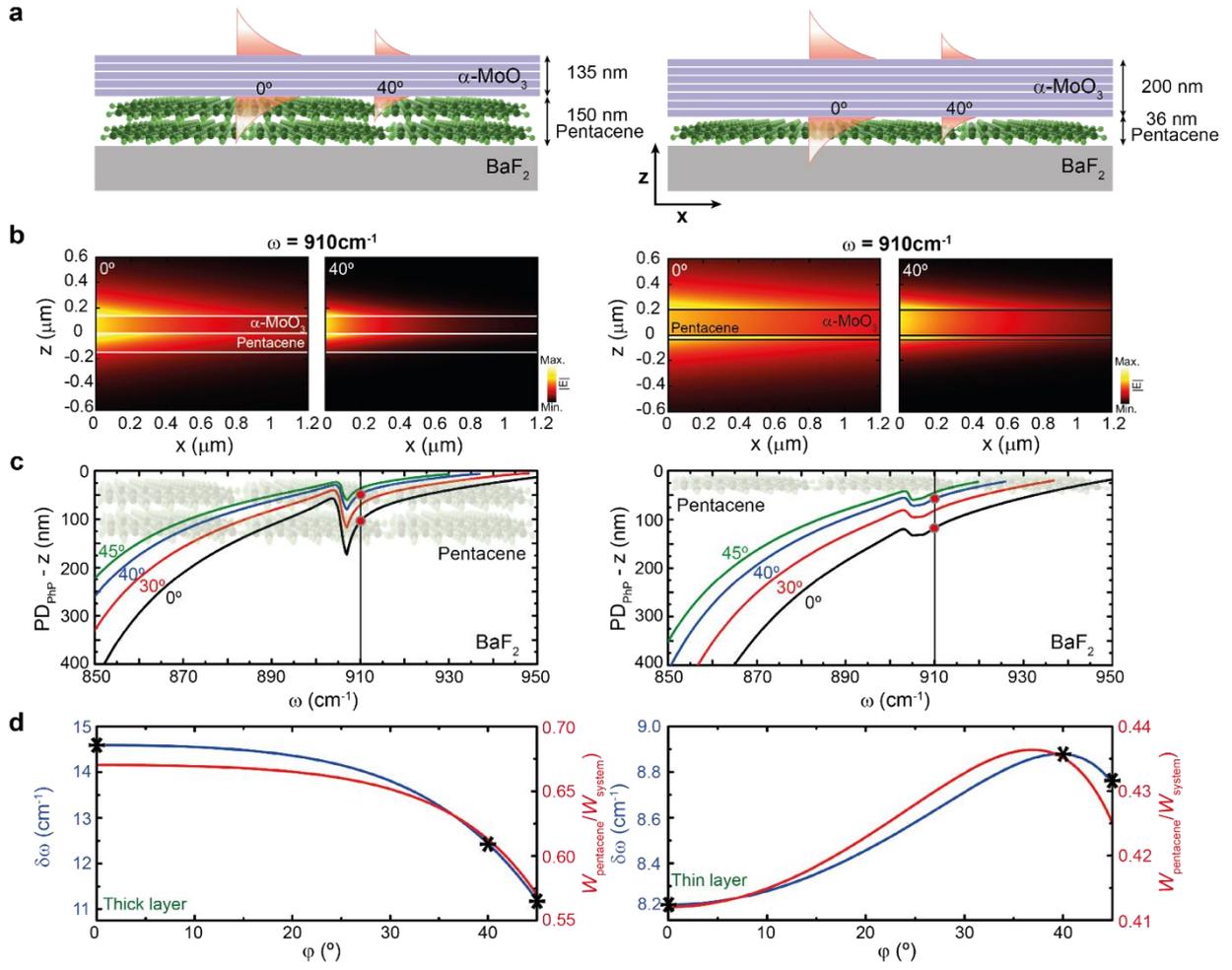

**Fig. 3. Analysis of directional VSC as a function of pentacene layer thickness. a**, Illustrations depicting the electromagnetic (EM) field cross-sections (x-z) at φ=0º and 40º to aid in understanding the different cases of having a thick pentacene layer (150 nm, where all the EM field is confined into the pentacene for all angles), and a thin pentacene layer (36 nm, where the EM field is predominantly confined into the pentacene for a large in-plane angle of 40º and high frequencies). **b**, Analytical calculations of the EM field (|E|) cross-sections (x-z) at 910 cm$^{-1}$ (representative case) for a thick (left) and a thin (right) pentacene layer, corresponding to φ=0º and 40º angles, corroborating the illustrations in (a). **c**, Left: Comparative analysis of the PhP field penetration depth (z-point where the norm of the polaritonic field decays by 1/e, PD$_{PhP}$ - z) from a 150 nm-thick pentacene layer to the substrate for four different in-plane angles: φ=0º (black curve), 30º (red), 40º (blue), and 45º (green). Right: The same for a 36 nm thick pentacene layer. The EM field only decays completely into the pentacene for the thick layer, suggesting a larger δω. **d**, Left: Comparison between the analytical calculation of δω (blue solid line) for a thick pentacene layer (curve shown in Fig. 2c) and the fraction of energy in the pentacene, relative to the total energy of the system ($W_{pentacene}/W_{system}$) at 910 cm$^{-1}$ (red solid line). Right: Same than on the left panel for a thin pentacene layer (curve shown in Fig. 2f). The decrease of $W_{pentacene}/W_{system}$ at large in-plane angles explains the similar decreasing tendency of δω for both thick and thin pentacene layers. The experimentally analyzed in-plane angles are indicated by black symbols.

To understand the underlying causes of the existence of this optimal condition and the different coupling trends observed for different pentacene thicknesses, we performed a theoretical study (Fig. 3) focusing on the overlap between the polaritonic electromagnetic (EM) fields and the pentacene layers as a function of the in-plane angle. We first calculated analytically the EM field ($|E|$) of the polaritonic fundamental mode (see Supplementary Note IV) along the x-z axes for the cases of $\varphi=0°$ and $40°$ (Fig. 3b). For these calculations, we chose a frequency of 910 cm$^{-1}$, i.e., slightly off from the molecular resonance absorption, allowing us to clearly visualize the PhP field. This frequency provides a representative example of the two branches, except for the frequencies close to the resonance frequency of the molecule, where almost no propagation of polaritons is expected due to the molecular resonance absorption.

For the thick pentacene layer (left column in Fig. 3), we observe that the field below the α-MoO$_3$ flake decays almost entirely within the pentacene layer for all in-plane angles (see left panel in Fig. 3b for the representative cases of $\varphi=0°$ and $40°$), indicating a condition of maximum overlap with the molecular vibrations (a schematic is shown in the left panel of Fig. 3a to better visualize this condition). This observation can be further corroborated by analysing the penetration depth (PD$_{PhP}$ - z) of the polaritonic fields into the pentacene layer (z-point where the norm of the polaritonic field decays by 1/e). The resulting plots (Fig. 3c) show that for frequencies close to the molecular resonance (904 cm$^{-1}$), the polaritonic field decays almost completely into the pentacene layer for all in-plane angles, indicating a maximum overlap (see Supplementary Note VI). Finally, to understand the trend followed by $\delta\omega$ as a function of in-plane angle (blue curve in Fig. 3d, shown previously in Fig. 2) we numerically calculated the fraction of energy in the pentacene layer relative to the total energy of the system $W_{pentacene}/W_{system}$ (red curve in Fig. 3d) at 910 cm$^{-1}$ (see Supplementary Note IV). We observe that it follows a very similar trend to $\delta\omega$, remaining almost constant until around $\varphi=30°$, where it starts to decay, showing a non-uniform dependence on the in-plane angle. This means that at larger in-plane angles the total polaritonic energy contained in the pentacene layer decreases, resulting in a decay of the total absorption of the molecules and thus explaining the behaviour observed for $\delta\omega$. Consequently, we can associate the trend of $W_{pentacene}/W_{system}$ with the polaritonic confinement within the biaxial slab for high in-plane angles[28].

Interestingly, for the thin pentacene layer (right column of Fig. 3), we observe a completely different behaviour. As shown in the cross-sections of the polaritonic field (Fig. 3b) and also illustrated at the top schemes (Fig. 3a), a significant part of the polaritonic field is located outside the pentacene layer. Notably, the fraction of the field that decays outside the pentacene layer (into the BaF$_2$ substrate) is reduced for larger in-plane angles. This can be clearly observed in the penetration depth representation (Fig. 3c), whose trend can be explained by the increase of the polaritonic confinement expected for large values of $\varphi$. Because the entire EM field is far from being confined within the pentacene, $\delta\omega$ takes smaller values than in the case of the thick pentacene layer (~8-9 cm$^{-1}$ compared to ~11-15 cm$^{-1}$). Note that this difference is significantly larger at smaller angles as we are farther from the maximum splitting obtained when the whole EM interacts with the pentacene molecules (as in the thicker pentacene layer case). To understand the different $\delta\omega$ trend observed in this case (red curve in Fig. 3d), we again calculate the total energy of the polaritonic field in the pentacene layer $W_{pentacene}/W_{system}$ (blue curve in Fig. 3d). A similar behaviour between both quantities is observed. The polaritonic energy in the pentacene layer gradually increases with the in-plane angle until it reaches a maximum at around $\varphi=40°$, after which it starts to decrease. We explain this behaviour as a combination of two differentiated effects. For small in-plane angles, the total polaritonic energy generated by the α-MoO$_3$ layer remains almost constant, as explained for the case of the thicker pentacene layer. Thus, increasing the percentage of the field that decays inside the thin pentacene layer results in an increase in the total energy contained in it, which is responsible for the δω increase. However, for higher in-plane angles, the maximum energy available decreases so quickly that this effect has more impact on the δω response than that from the increase of the confinement, resulting in an overall decrease of δω. When both effects compensate, we obtain a maximum in the total energy representation, which is also observed in the δω plot. The difference between the two maximums is related to the chosen frequency for the total energy calculation. Lower energies, such as 900 cm$^{-1}$, exhibit lower maximum values as a result of the higher penetration depths (see Supplementary Note IV). Overall, we can state that δω shows a maximum coupling at one specific direction, corresponding to the threshold at which the increase in the fraction of the field absorbed by the pentacene layer is not compensated by the decrease of the overall polaritonic intensity outside the α-MoO$_3$ slab at large angles. This result unambiguously reveals a condition of a unique direction providing an optimum in directional VSC phenomena. Importantly, for a given molecular layer the optimal direction can be adjusted by simply rotating the polaritonic layer (as schematically shown in the inset of Fig. 1a) as well as choosing its most convenient thickness (note that for PhPs in a thin hyperbolic layer the polariton wavelength, and thus penetration depth, directly depends on the thickness).

In summary, our findings introduce the concept of directionality in the phenomenon of SC by demonstrating that the coupling strength between PhPs in the strongly anisotropic vdW crystal α-MoO$_3$ and molecular vibrations in pentacene exhibit a clear directional dependence. Moreover, when the molecular layer is sufficiently thin, this directional VSC phenomenon can exhibit an optimum characterized by a maximal light-matter coupling along one specific direction. These results open the door to potential applications for the detection, manipulation, or control of chemical properties[3-7] along specific directions at the nanoscale, such as those where a specific target bond is present (directional nano-chemistry).

## Methods

**Thermal evaporation of pentacene.** Pentacene molecules of sublimed quality (99.9%) (Sigma-Aldrich, Saint Louis, MO, USA) were thermally evaporated in an ultra-high vacuum evaporator chamber (base pressure ~$10^{-9}$ mbar), at a temperature of 170ºC and a rate of 0.2 nm s$^{-1}$ using a Knudsen cell. To prevent the formation of rough polycrystalline films and obtain smoother amorphous films, the substrates were placed onto a cold finger which was cooled down to a temperature of approximately 80 K by liquid nitrogen.

**α-MoO$_3$ sample preparation.** Bulk α-MoO$_3$ crystals were mechanically exfoliated using Nitto tape (Nitto Denko Co., SPV 224P). Subsequently, a second exfoliation was performed from the tape onto transparent polydimethylsiloxane (PDMS) to reduce their thickness. The flakes were examined under an optical microscope to select uniform pieces with the desired thicknesses (approximately 100-200 nm) and large surface areas. The flakes were then transferred onto the substrate with pentacene that had been thermally evaporated using the dry transfer technique.

**Fourier-Transform Infrared Spectroscopy.** Fourier-Transform Infrared Spectroscopy (FTIR) was employed to characterize the optical response of pentacene in the far-field and determine its permittivity. For this purpose, a Varian 620-IR microscope connected to a Varian 670-IR spectrometer was utilized, along with a broadband mercury cadmium telluride (MCT) detector operating within the range of 400-6000 cm$^{-1}$. The reflectance spectra were collected with a spectral resolution of 2 cm$^{-1}$. The IR radiation emitted by the thermal source was incident at a normal angle. BaF$_2$ substrates were selected due to their transparency within the studied spectral range. An Au layer was employed as a reference for normalization.

**Scattering-Scanning Near Field Optical Microscopy.** Near-field imaging measurements were conducted using a commercially available scattering-type Scanning Near Field Optical Microscope (s-SNOM) provided by Neaspec GmbH. The microscope was equipped with a quantum cascade laser from Daylight Solutions (890-1140 cm$^{-1}$). Metal-coated (Pt/Ir) atomic force microscopy (AFM) tips (ARROW-NCPt-50, Nanoworld) operating at a tapping frequency $\Omega \sim 280$ kHz and an oscillation amplitude of approximately 100 nm were utilized as the source and probe for detecting polaritonic excitations. The light scattered by the tip was focused onto an IR detector (Kolmar Technologies) using a parabolic mirror. To suppress background noise, demodulation of the detected signals to the 3rd harmonic of the tip frequency ($s_3$) was performed. A pseudo-heterodyne interferometric method was employed to extract both amplitude and phase signals independently.

**Eigenmode analysis.** The dispersion relation of hybridized molecular-PhP modes with complex wave vectors ($k+i\kappa''$) was obtained through both numerical simulations using COMSOL (software based on the finite-element method in the frequency domain), and analytical calculations based on the boundary conditions of Maxwell's equations in the large momentum approximation, to ensure consistency and accuracy. As an initial step, the value of the wave vector $k$ obtained from analytical calculations was used in the standard iterative mode search module of COMSOL, leading to the identification of the region with high losses in the dispersion curve (known as the back-banding region). Analytically, using the complex frequency approach ($\omega-i\gamma$), two separate branches of dispersion and the mode splitting, $\delta\omega$, were identified. The distribution of the electric field magnitude, $|E|$, was calculated analytically, depending on the flake angle, $\theta$, and the in-plane coordinate, $x$. The percentage of the electromagnetic field energy within the pentacene layers, which explains the dependence of the mode splitting, $\delta\omega$, with the flake angle, $\varphi$, was calculated using the equation for electromagnetic energy in media with dispersion.

## Acknowledgments
The authors acknowledge support from the Spanish Ministry of Science and Innovation (grants PID2022-141304NB-I00, and PID2020-115221GB-C42). A.I.F.T.-M. and J.A.-C. acknowledge support through the Severo Ochoa program from the Government of the Principality of Asturias (nos. PA-21-PF-BP20-117 and PA-22-PF-BP21-100). K.V.V. acknowledges support from a fellowship from "la Caixa" Foundation (ID 100010434), the fellowship code is LCF/BQ/DI21/11860026. J.T.-G. acknowledges support from the Swiss National Foundation (grant no. 200020_201096). A.M. and P.D.-N. acknowledge support from the European Research Council (ERC) under the European Union's Horizon 2020 research and innovation program (Consolidator Grant Agreement No. 865590, Programable Matter). J.M.-S. acknowledges financial support through the Ramón y Cajal program from the Government of Spain (RYC2018-026196-I). P.A.-G. and the European Research Council under Consolidator grant No. 101044461, TWISTOPTICS. A.Y.N. acknowledges the Basque Department of Education (grant PIBA-2023-1-0007).

## Author contributions
P.A.-G. and A.Y.N. conceived the study. A.I.F.T.-M. performed the sample fabrication and near-field experiments with the help of J.T.-G., A.T.M.-L., and J.D. O.G.M. performed the analytical study and simulations with input from C.L., J.A.-C., and K.V.V. and supervised by V.S.V. and A.Y.N. P.D.-N. and A.M. performed the nano-FTIR measurements. F.C. performed the fabrication of the pentacene films under the supervision of L.E.H. A.B. and R.H. participated in the data analysis. A.I.F.T.-M. and P.A.-G. wrote the manuscript with input from the rest of the authors. P.A.-G. and A.Y.N. supervised the work. All authors contributed to the scientific discussion and manuscript revisions.

## Competing interests
R.H. is a co-founder of Neaspec GmbH, a company that manufactures scattering-type scanning near-field optical microscope systems, including the one used in this study. The other authors declare no conflicts of interest.

Supplementary information for:

**"Directional strong coupling at the nanoscale between hyperbolic polaritons and organic molecules"**

**Supplementary sections:**

**Supplementary Note I. In-plane polariton anisotropy from near-field imaging at different frequencies on thin and thick pentacene layers.**

**Supplementary Note II. Analysis of the strong coupling.**

**Supplementary Note III. PhP propagation length at the region of anomalous dispersion (back-bending).**

**Supplementary Note IV. Analysis of the hybrid molecular-polaritonic modes for complex wave vectors or complex frequencies.**

**Supplementary Note V. Dielectric function and growth parameters of pentacene.**

**Supplementary Note VI. Analytical study of the polaritonic fields.**

# Supplementary Note I. In-plane polariton anisotropy from near-field imaging at different frequencies on thin and thick pentacene layers

To extract the experimental data shown in the dispersions in Fig. 2 of the main text, numerous s-SNOM near-field amplitude ($s_3$) measurements were carried out at different frequencies for each in-plane angle. Supplementary Figs. 1 and 2 show examples of the near-field measurements conducted for an in-plane angle φ=0° for a 36 nm thick (Supplementary Fig. 1) and a 150 nm thick (Supplementary Fig. 2) pentacene layer. To extract the polariton wavelength ($\lambda_p$) we take a profile (shown as a gray solid line in first image of Supplementary Figs. 1a and 2a, and similarly for the rest of the images below) as shown in Supplementary Figs. 1b and 2b. From these profiles we extract $\lambda_p$ for each frequency (red dashed lines) by fitting them using the expression:

$$A + \frac{B \cdot e^{-\frac{x}{t_0}} \cdot \sin\left(\frac{2\pi(x-x_{c0})}{\lambda}\right)}{x} + \frac{C \cdot e^{-2\frac{x}{t_0}} \cdot \sin\left(\frac{4\pi(x-x_{c1})}{\lambda}\right)}{\sqrt{x}} \qquad (S1)$$

which combines two decaying oscillatory terms, each with different decay rates and oscillation frequencies, corresponding to the tip and edge contributions to the polaritonic field, respectively. The profiles shown in Supplementary Figs. 1b and 2b exhibit both contributions as doublets[1]. Interestingly, the PhP wavelength changes its tendency at the back bending region, which constitutes a fingerprint of strong coupling for propagating polaritons.

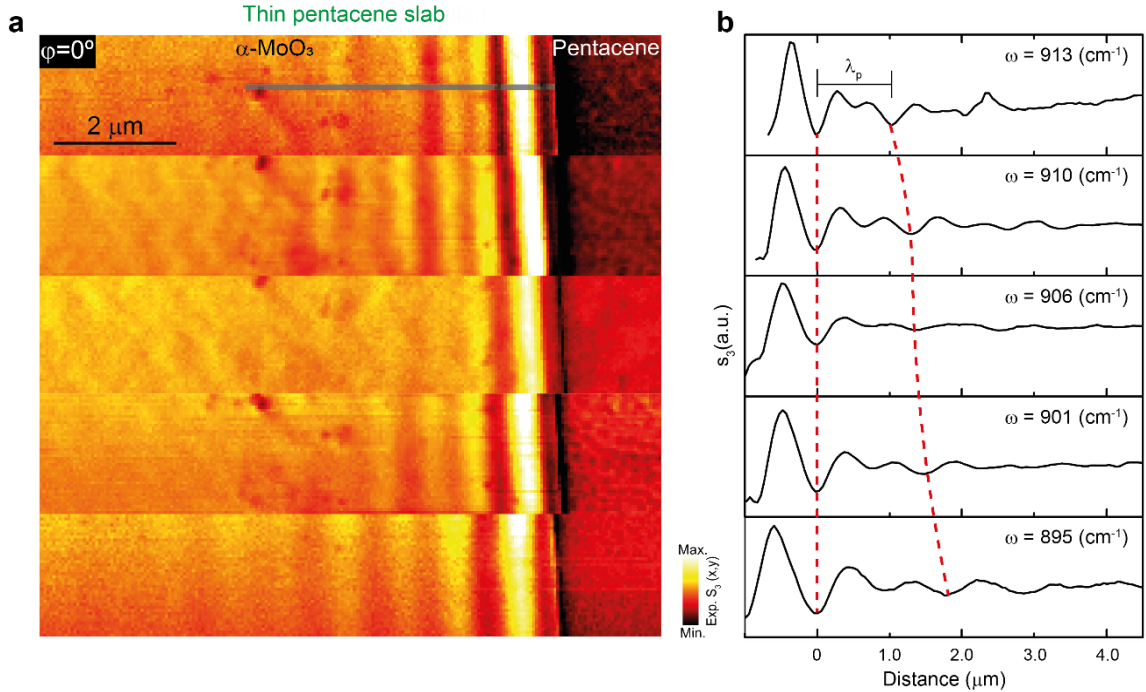

**Supplementary Fig. 1. Near-field s-SNOM imaging of PhPs in a flake of α-MoO₃ (200nm) on top of a thin (36nm) pentacene layer at different frequencies. a**, Near-field amplitude s-SNOM images ($s_3$) at different frequencies for an in-plane angle of 0°. **b**, Profiles extracted from the gray solid line in **a** for each image at a different frequency.

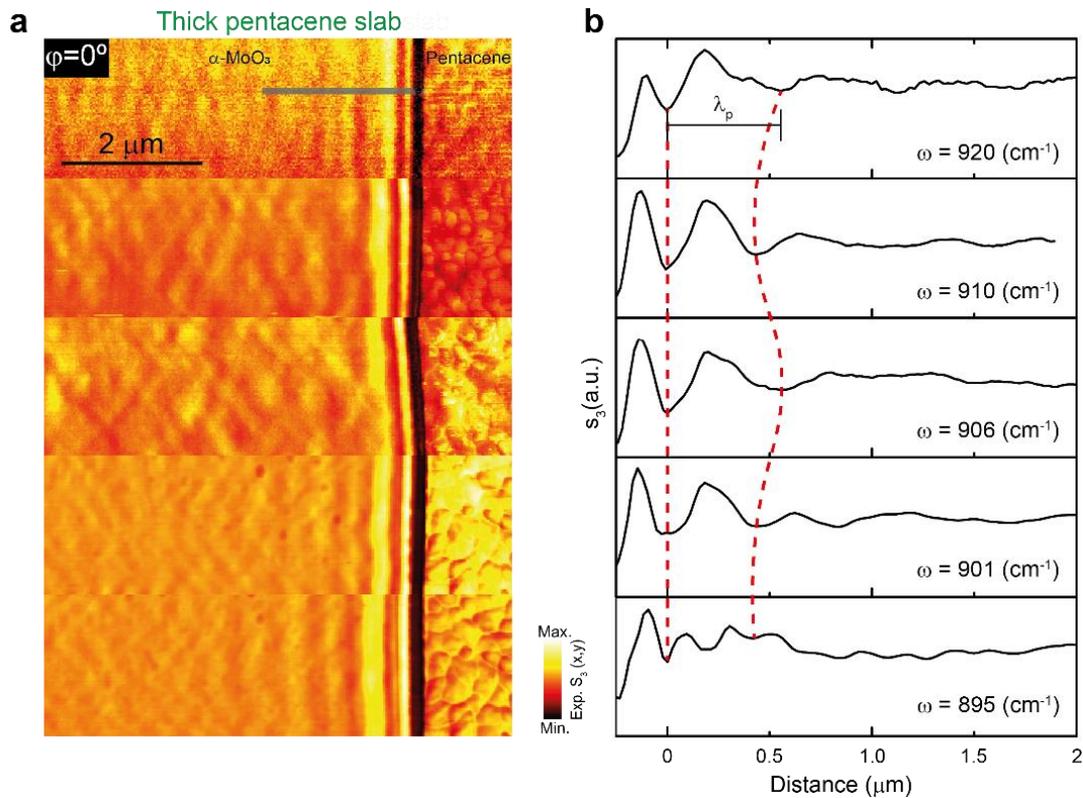

**Supplementary Fig. 2. Near-field s-SNOM imaging of PhPs in a flake of α-MoO$_3$ (135nm) on top of a thick (150nm) pentacene layer at different frequencies. a**, Near-field amplitude s-SNOM images ($s_3$) at different frequencies for an in-plane angle φ=0º. **b**, Profiles extracted from the gray solid line in **a** for each image at a different frequency. The polariton wavelength ($\lambda_p$) is delimited by a black solid line for the first image. Red dashed lines illustrate the anomalous evolution in $\lambda_p$ at different frequencies.

When the number of fringes is limited, such as in the back-bending region or the upper branch, it is also possible to extract the polariton wavelength by performing a Fourier Transform (FT) at the center of the s-SNOM images ($s_3$), as shown in Supplementary Fig. 3a. Due to the roughness of the molecular layer, PhPs disperse and form specific patterns in the s-SNOM images (Supplementary Figs. 1a, 2a, and 3a). By performing the FT in this zone (Supplementary Fig. 3a), we obtain the hyperbolic IFC for the specific frequency (Supplementary Fig. 3b), from which we can extract the polariton wavelength for the most challenging cases.

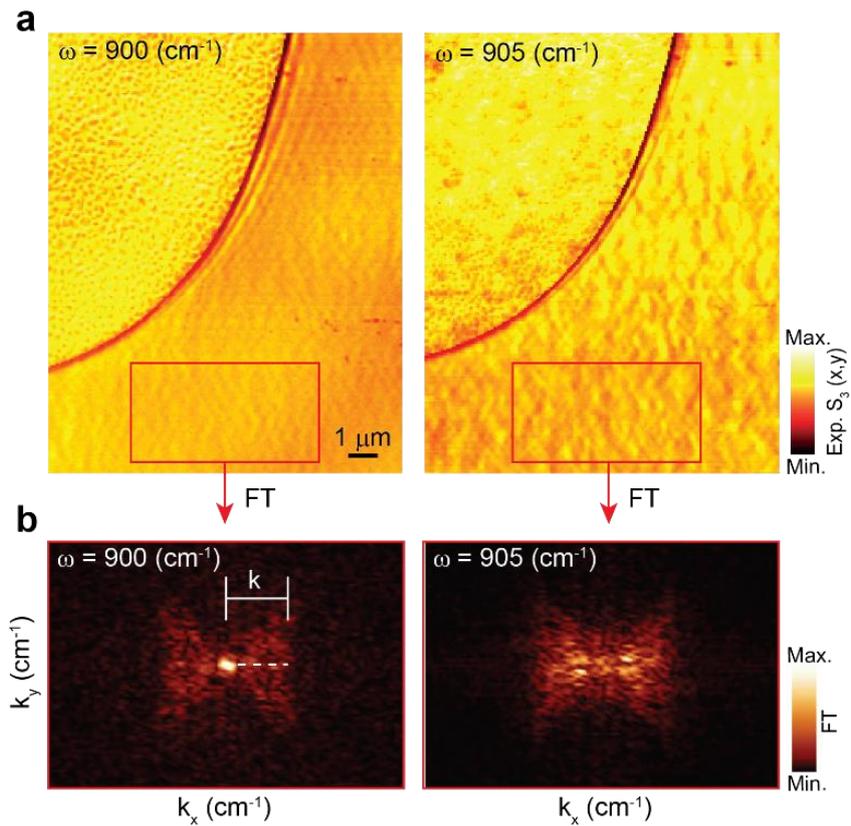

**Supplementary Fig. 3. Fourier transforming of central areas from near-field imaging with s-SNOM. a,** Near-field amplitude s-SNOM images ($s_3$) at frequencies $\omega = 900$ cm$^{-1}$ and $\omega = 905$ cm$^{-1}$ for a curved flake. **b,** Isofrequency curve (IFC) obtained by Fourier transforming (FT) the real-space measurements from the areas inside the red square in **a**.

## Supplementary Note II. Analysis of the strong coupling

To quantify if the interaction between PhPs and molecular vibrations of pentacene reaches the SC condition, we use the mathematical expression[2]:

$$C \stackrel{\text{def}}{=} \frac{\delta\omega^2}{\frac{\Gamma_{PhP}^2}{2}+\frac{\Gamma_{mol}^2}{2}} > 1 \qquad (S2)$$

where $\delta\omega$ is the Rabi splitting between the PhP dispersion branches, while $\Gamma_{PhP}$ and $\Gamma_{mol}$ are the uncoupled PhP and molecule linewidths, respectively. The Rabi splitting is calculated at the momentum value where the uncoupled PhP dispersion meets the molecular resonance (Supplementary Fig. 4). This resonance occurs at 904 cm$^{-1}$, and its linewidth is found to be $\Gamma_{mol} = 3.5$ cm$^{-1}$ (see Supplementary Note V). On the other hand, the uncoupled system consists of an α-MoO$_3$ layer with a dielectric spacer below, optically characterised by the high-frequency value of the pentacene permittivity ($\varepsilon_\infty = 2.98$) and placed on top of a BaF$_2$ substrate. Note that, due to the in-plane anisotropy of α-MoO$_3$ PhPs, the polariton momentum will vary with the in-plane direction. The dispersion of PhPs is calculated using an analytical model that considers complex-valued momenta and real frequency (see Supplementary Note IV).

After determining the momentum value for computing the Rabi splitting, we fixed the real part of this value and used the same analytical model, but with complex frequency ($\omega - i\frac{\Gamma_{PhP}}{2}$), to obtain the PhP linewidth $\Gamma_{PhP}$, yielding 3.8 cm$^{-1}$ for φ = 0° and $\Gamma_{PhP} = 4.6$ cm$^{-1}$ for φ = 40°.

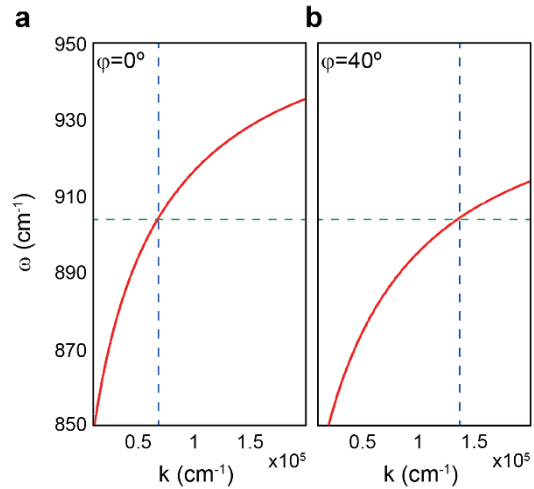

**Supplementary Fig. 4. Theoretical dispersions for the uncoupled system. a**, Analytical dispersions (red solid line) considering complex-valued momenta and real frequency of PhPs in α-MoO$_3$ (135 nm in thickness) with a dielectric spacer of 150 nm in thickness, optically characterised by the high-frequency value of the pentacene permittivity ($\varepsilon_\infty = 2.98$), and placed on top of a BaF$_2$ substrate for φ = 0°. **b**, Same as (a) for φ = 40°. The green dashed line indicates the frequency of absorption of the molecule ($\omega = 904$ cm$^{-1}$), while the blue dashed line indicates the point at the momentum (k) where the uncoupled PhP dispersion meets the molecular resonance.

# Supplementary Note III. PhP propagation length at the region of anomalous dispersion (back-bending)

In Supplementary Fig. 5, we show s-SNOM near-field images of curved α-MoO$_3$ flakes on top of a 36 nm thick (Supplementary Fig. 5a) and a 150 nm thick (Supplementary Fig. 5b) pentacene layer, at four different frequencies. We can observe the same evolution noted previously at φ=0° in Supplementary Note I but, in this case, for any in-plane direction. Upon reaching the absorption frequency of the molecule (904 cm$^{-1}$), the propagation length (L$_{PhP}$) decreases.

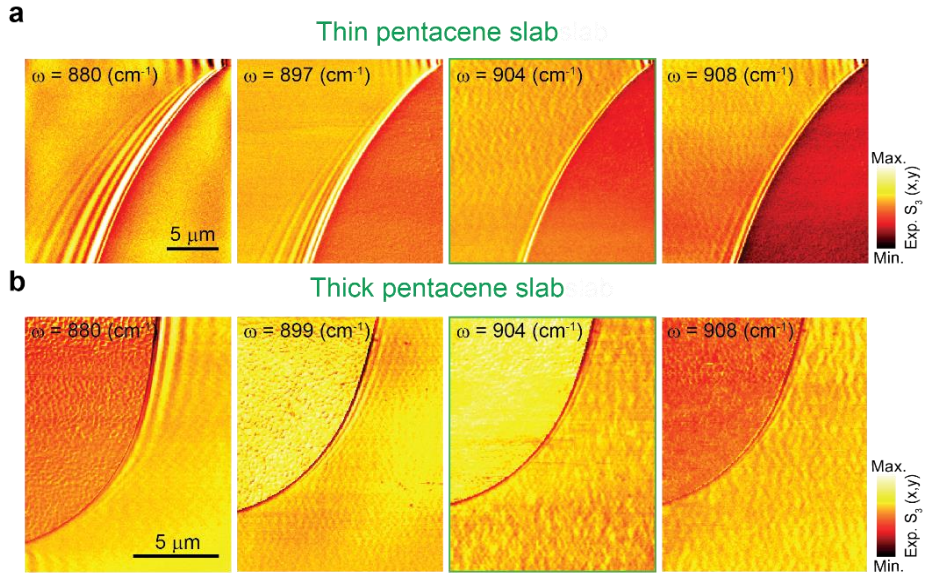

**Supplementary Fig. 5. Near-field s-SNOM imaging of PhPs in a curved α-MoO$_3$ flake on top of pentacene layers of different thicknesses. a**, Near field amplitude (s$_3$) for the case of a thin pentacene slab (36 nm) at four different frequencies, $\omega = 880, 897, 904,$ and $908$ cm$^{-1}$, showing the polaritonic evolution passing through the absorption frequency of the molecule. **b**, Near field amplitude (s$_3$) for the case of a thick pentacene slab (150 nm) at four different frequencies, $\omega = 880, 899, 904,$ and $908$ cm$^{-1}$, showing the polaritonic evolution passing through the absorption frequency of the molecule.

This dramatic change in L$_{PhP}$ is clearly illustrated in Supplementary Figs. 6a and 6b where we plot the analytically calculated PhPs propagation length L$_{PhP}$ (1/Im(k)) as a function of frequency for both coupled and uncoupled systems (for two cases of thick and thin pentacene layers). The anomalous propagation, resulting from the absorption of PhPs at the vibrational frequency of the pentacene molecule (almost no propagation is observed in the experimental images and confirmed through the significantly reduced propagation lengths for the coupled systems compared to the uncoupled systems), introduces the back-bending effect observed in the dispersions of Fig. 2 of the main text, indicative of SC.

Additionally, in Supplementary Fig. 5, we observe an increasing momenta and shorter wavelengths with increasing frequency in the upper branch.

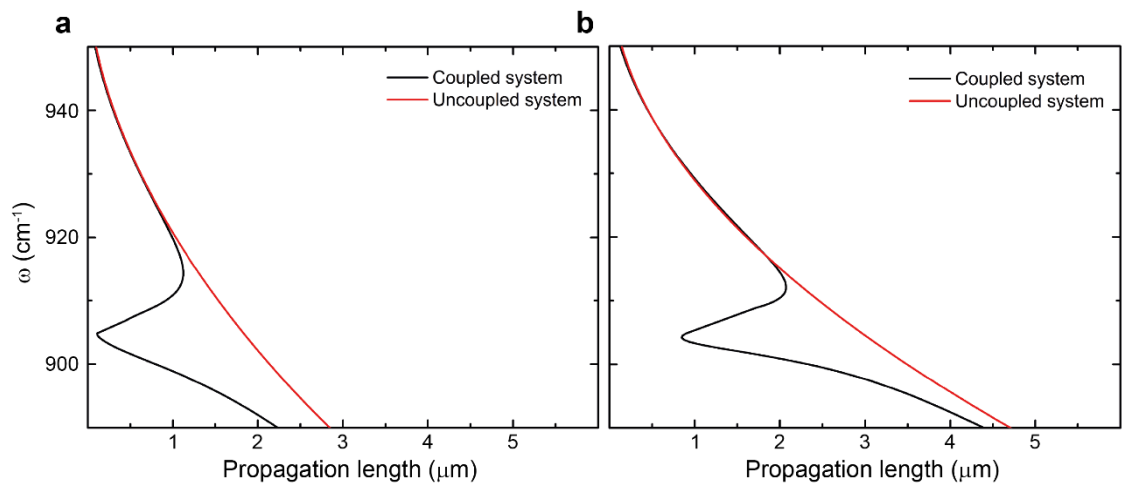

**Supplementary Fig. 6. Propagation lengths for the coupled and uncoupled systems.**
**a**, Thick pentacene layer (150 nm). **b**, Thin pentacene layer (36 nm).

# Supplementary Note IV. Analysis of the hybrid molecular-polaritonic modes for complex wave vectors or complex frequencies.

The analytical tools used throughout this work are summarized in this section. As a first step, we have obtained the dispersion relation of polaritons in a 4-media heterostructure under the high-momentum approximation[3]. Such a heterostructure is shown in Supplementary Fig. 7.

The electric fields in each layer are shown next:

$$E_0 = a_0 \begin{pmatrix} q_x \\ q_y \\ iq \end{pmatrix} e^{ik_x x + ik_y y} e^{-q k_0 z}, z > d \tag{S3}$$

$$E_1 = a_1 \begin{pmatrix} q_x \\ q_y \\ q_{z1} \end{pmatrix} e^{ik_x x + ik_y y} e^{i q_{z1} k_0 z} + b_1 \begin{pmatrix} q_x \\ q_y \\ -q_{z1} \end{pmatrix} e^{ik_x x + ik_y y} e^{-i q_{z1} k_0 z}, 0 < z < d \tag{S4}$$

$$E_2 = a_2 \begin{pmatrix} q_x \\ q_y \\ iq \end{pmatrix} e^{ik_x x + ik_y y} e^{-q k_0 z} + b_2 \begin{pmatrix} q_x \\ q_y \\ -iq \end{pmatrix} e^{ik_x x + ik_y y} e^{q k_0 z}, -t < z < 0 \tag{S5}$$

$$E_3 = b_0 \begin{pmatrix} q_x \\ q_y \\ -iq \end{pmatrix} e^{ik_x x + ik_y y} e^{q k_0 z}, z < -t \tag{S6}$$

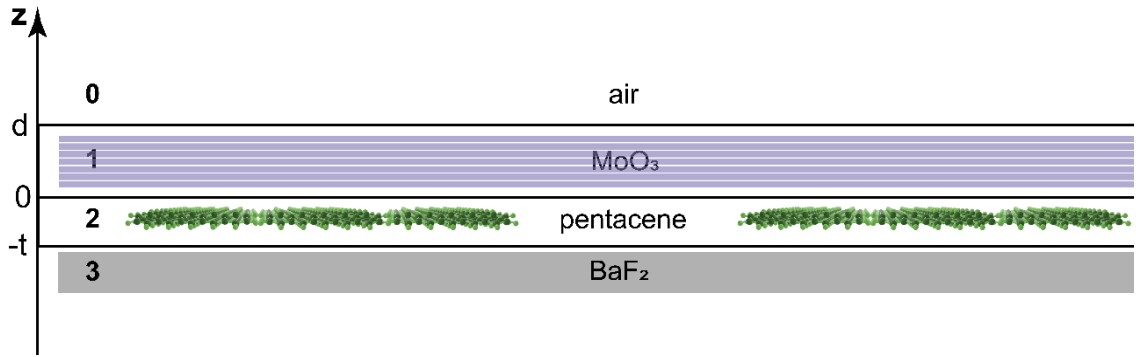

**Supplementary Fig. 7. Sketch of the 4-media structure.**

To obtain the polariton dispersion relation, we compute the boundary conditions for the in-plane components of **E** and **H** at the boundaries air/α-MoO$_3$ ($z = d$), α-

MoO$_3$/pentacene ($z = 0$) and pentacene/BaF$_2$ (z=-t). For **E,** we have the following equations:

$$\begin{aligned} \mathbf{E}_{0t}(z = d) &= \mathbf{E}_{1t}(z = d) \\ \mathbf{E}_{1t}(z = 0) &= \mathbf{E}_{2t}(z = 0) \\ \mathbf{E}_{2t}(z = -t) &= \mathbf{E}_{3t}(z = -t) \end{aligned} \quad (S7)$$

Similarly, the boundary conditions for the in-plane components of **H** are the following:

$$\begin{aligned} \mathbf{H}_{0t}(z = d) &= \mathbf{H}_{1t}(z = d) \\ \mathbf{H}_{1t}(z = 0) &= \mathbf{H}_{2t}(z = 0) \\ \mathbf{H}_{2t}(z = -t) &= \mathbf{H}_{3t}(z = -t) \end{aligned} \quad (S8)$$

Where we have used the relation:

$$\mathbf{H}_i = \mathbf{q}_i \times \mathbf{E}_i \quad (S9)$$

Eqs. S7 and S8 can be written in the matrix form with help of Eq. S9:

$$\begin{pmatrix} -A & B & B^{-1} & 0 & 0 & 0 \\ i\varepsilon_1 qA & -\varepsilon_z q_z B & \varepsilon_z q_z B^{-1} & 0 & 0 & 0 \\ 0 & 1 & 1 & -1 & -1 & 0 \\ 0 & \varepsilon_z q_z & -\varepsilon_z q_z & -i\varepsilon_p q & i\varepsilon_p q & 0 \\ 0 & 0 & 0 & C & C^{-1} & C^{-1} \\ 0 & 0 & 0 & i\varepsilon_p qC & -i\varepsilon_p qC & i\varepsilon_{BaF_2} qC^{-1} \end{pmatrix} \begin{pmatrix} a_0 \\ a_1 \\ b_1 \\ a_2 \\ b_2 \\ b_0 \end{pmatrix} = \vec{0}, \quad (S10)$$

where $A = e^{-qk_0 d}, B = e^{iq_z k_0 d}, C = e^{qk_0 t}$. To find a non-trivial solution of the system S10, the determinant of the 6x6 matrix should be equal to 0. This gives us the following equation, which constitutes the polariton dispersion of the system under the high-momentum approximation:

$$\varepsilon_p q q_z \varepsilon_z (\varepsilon_{BaF_2} + \varepsilon_1) + \tan(q_z k_0 d) \varepsilon_p [\varepsilon_{BaF_2} q^2 \varepsilon_1 - \varepsilon_z^2 q_z^2] + \\ \tanh(qk_0 t) q q_z \varepsilon_z [\varepsilon_p^2 + \varepsilon_{BaF_2} \varepsilon_1] + \tan(q_z k_0 d) \tanh(qk_0 t) [\varepsilon_p^2 q^2 \varepsilon_1 - \varepsilon_{BaF_2} \varepsilon_z^2 q_z^2] = 0 \quad (S11)$$

$$q = \frac{k}{2\pi\omega}, \quad q_z = \frac{k}{2\pi\omega} \frac{\sqrt{\frac{\varepsilon_x}{\varepsilon_z} \cos^2\varphi + \frac{\varepsilon_y}{\varepsilon_z} \sin^2\varphi}}{i} \quad (S12)$$

where $\omega$ is the frequency of incident light, $\varphi$ is the in-plane angle of the anisotropic crystal between the [100] axis and the direction of the polariton wave propagation vector **k,** and $k = |\mathbf{k}|$. Expression S10 gives the dependence $q(\omega)$ in implicit form, so the search for solutions will be carried out using a numerical method implemented in Matlab. This search can rely on two strategies: using real-valued frequencies and complex-valued momenta or complex-valued frequencies and real-valued momenta.

For the real-valued frequencies and complex-valued momenta, let we set:

$$k = \mathrm{k} + i\kappa$$
$$\omega \in \mathbb{R} \quad \text{(S13)}$$

and numerically find the solution for hybrid modes in such a case.

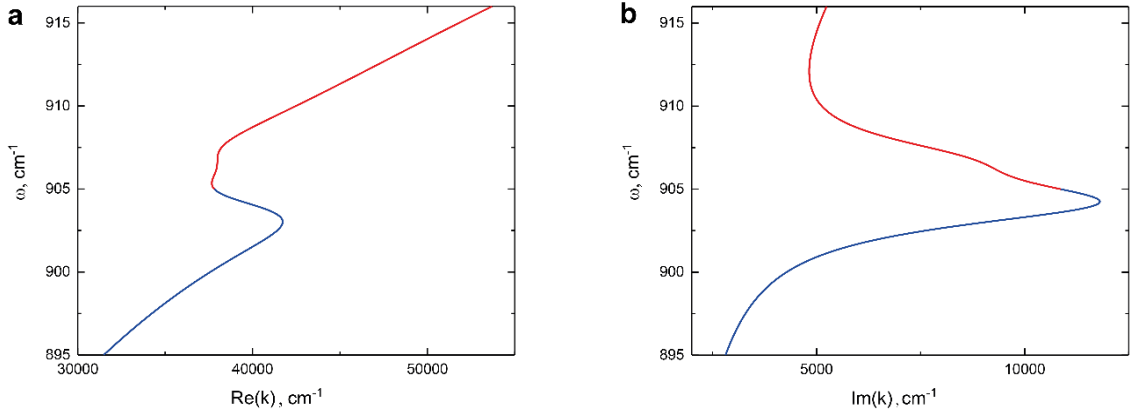

**Supplementary Fig. 8. Complex-k approach. a**, Frequency $\omega$ vs $\mathrm{Re}(k)$ dependence. **b**, Frequency $\omega$ vs $\mathrm{Im}(k)$ dependence. $\varphi = 0°$ for both panels **a**, **b**.

As an example of the solution of Eq. S13 for complex wavevector and real frequency, we show in Supplementary Fig. 8 the polariton dispersion for the system with $t = 36$ nm and $d = 200$ nm, at $\varphi = 0°$. The case of a complex wave vector and a real frequency is realized in experiment, so the region of the back-bending can be experimentally observed. However, in this approach we do not see the anti-crossing region of the hybrid mode branches. In order to observe it, we solve Eq. S14 using complex-valued frequencies and real-valued wavevectors:

$$\omega = \omega + i\gamma,$$
$$k \in \mathbb{R} \quad \text{(S14)}$$

and numerically find the solution for hybrid modes in such a case. For the system with $t = 36$ nm and $d = 200$ nm, the solution for $\varphi = 0°$ is shown in Supplementary Fig. 9.

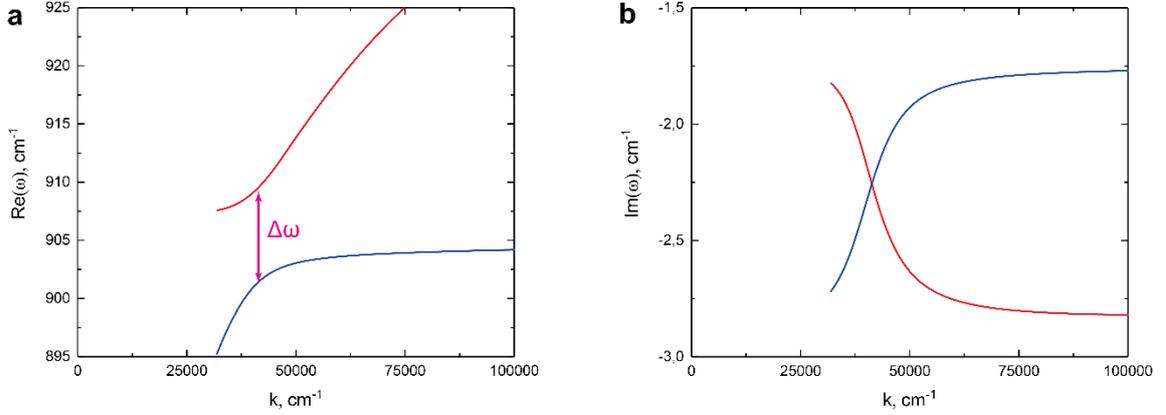

**Supplementary Fig. 9. Complex-ω approach**. **a**, Real part of frequency Re($\omega$) vs $k$ dependence. The pink arrow represents the mode splitting value, $\Delta\omega \equiv \delta\omega$ (the minimal distance between branches). **b**, Imaginary part of frequency Im($\omega$) vs. $k$ dependence. $\varphi = 0°$ for both panels **a, b**.

The strength of the strong coupling can be determined through the value of the mode splitting $\Delta\omega \equiv \delta\omega$ in the case of the complex frequency approach, as shown in Supplementary Fig. 10.

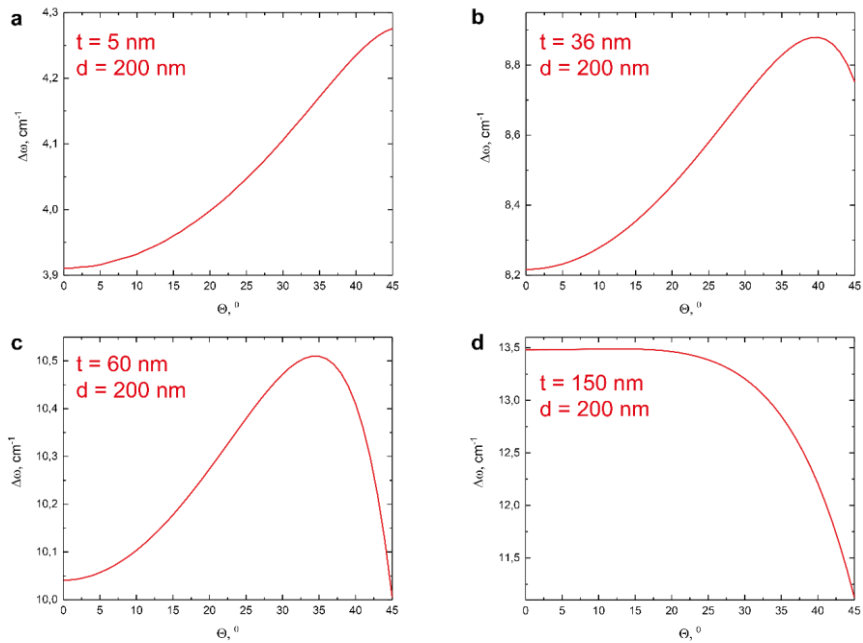

**Supplementary Fig. 10. Dependence of the mode splitting value *Δω* on the in-plane angle Θ ≡ *φ* for different pentacene layer thicknesses *t* and the α-MoO₃ thickness *d* = 200 nm**. **a**, $t = 5$ nm. **b**, $t = 36$ nm. **c**, $t = 60$ nm. **d**, $t = 150$ nm.

Supplementary Fig. 10 demonstrates the changing character of the $\Delta\omega$ vs. $\varphi$ dependency: for thin layers of the pentacene molecules, the strength of the SC will monotonically increase with increasing $\varphi$, whereas for thick layers, it will monotonically decrease. Consequently, for medium values of $t$, there is a maximum in the strength of the SC, as seen in Supplementary Figs. 10b and 10c.

**Dependence of the strong coupling strength on the in-plane angle for different pentacene layer thicknesses, expressed in terms of the energy stored in the pentacene layer relative to the total energy in the system:**

An equivalent approach to understand the directional strong coupling phenomenon relies on the study of the energy density distribution. In general, the energy density in a dispersing medium is written as follows[4]:

$$W = \frac{1}{16\pi}\left(\mathbf{E}^\dagger \frac{d(\omega\hat{\varepsilon})}{d\omega}\mathbf{E} + \mathbf{H}^\dagger \frac{d(\omega\hat{\mu})}{d\omega}\mathbf{H}\right) \quad (S15)$$

Where:

$$\hat{\varepsilon} = \begin{pmatrix} \varepsilon_x & 0 & 0 \\ 0 & \varepsilon_y & 0 \\ 0 & 0 & \varepsilon_z \end{pmatrix} \quad (S16)$$

is the permittivity tensor of an anisotropic medium. After computed Eq. S15 in each of the 4 layers of the system, we can calculate the fraction of energy in the pentacene relative to the total energy in the system $W_{pentacene}/W_{system}$.

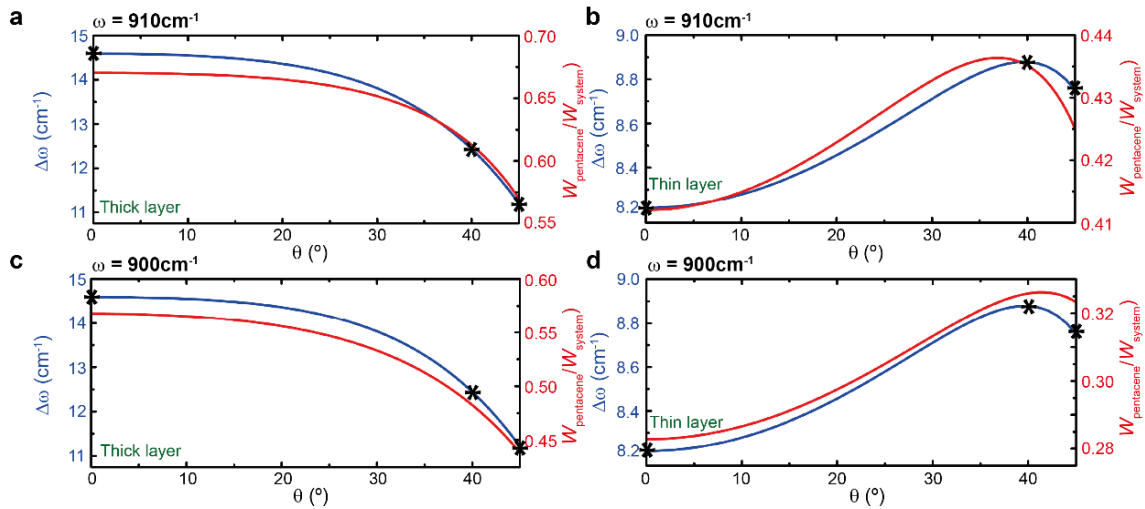

**Supplementary Fig. 11.** Visual comparison of the dependencies of the mode splitting value $\Delta\omega$, and the energy contribution of the pentacene layer relative to the total energy in the system $W_{pentacene}/W_{system}$ as a function of the in-plane angle $\theta \equiv \varphi$ for two systems and two different frequencies. **a**, Dependency

$W_{pentacene}/W_{system}$ on $\theta$ for $t = 150$ nm and $d = 135$ nm, compared with the dependency of $\Delta\omega$ on $\theta$ for $t = 150$ nm and $d = 135$ at 910 cm$^{-1}$. **b**, Dependency $W_{pentacene}/W_{system}$ on $\theta$ for $t = 36$ nm and $d = 200$ nm, compared with the dependency of $\Delta\omega$ on $\theta$ for $t = 36$ nm and $d = 200$ at 910 cm$^{-1}$. **c**, Dependency $W_{pentacene}/W_{system}$ on $\theta$ for $t = 150$ nm and $d = 135$ nm, compared with the dependency of $\Delta\omega$ on $\theta$ for $t = 150$ nm and $d = 135$ at 900 cm$^{-1}$. **d**, Dependency $W_{pentacene}/W_{system}$ on $\theta$ for $t = 36$ nm and $d = 200$ nm, compared with the dependency of $\Delta\omega$ on $\theta$ for $t = 36$ nm and $d = 200$ at 900 cm$^{-1}$.

For all panels in Supplementary Fig. 11, we computed $W$ for two different frequencies, 900 and 910 cm$^{-1}$, on the lower and upper branches. In Supplementary Fig. 11b at 910 cm$^{-1}$, it can be noticed that the $W_{pentacene}/W_{system}$ curve appears shifted to the left in comparison with the curve for $\Delta\omega$. On the other hand, In Supplementary Fig. 11d at 900 cm$^{-1}$ appears shifted to the right. Although we are able to observe the tendency of the maximum, it is not possible to calculate it at 904 cm$^{-1}$, where the absorption of the molecule occurs, since at this point all the field is absorbed by the molecule. However, these results allow us to understand that the exact tendency should be something in between, calculated around the absorption frequency.

Finally, we obtained the field amplitudes in Eq. S10 and derived expressions for the field amplitudes:

$$a_0 = 1$$

$$a_1 = \frac{A\,(\varepsilon_z q_z + 1i\varepsilon_1 q)}{2B\varepsilon_z q_z}$$

$$b_1 = \frac{A\,(\varepsilon_z q_z - 1i\varepsilon_1 q)}{2B^{-1}\varepsilon_z q_z}$$

$$a_2 = \frac{iq\varepsilon_p(a_1 + b_1) + \varepsilon_z q_z(a_1 - b_1)}{2iq\varepsilon_p}$$

$$b_2 = \frac{iq\varepsilon_p(a_1 + b_1) - \varepsilon_z q_z(a_1 - b_1)}{2iq\varepsilon_p}$$

$$b_0 = \frac{Ca_2 + C^{-1}b_2}{C^{-1}} \tag{S17}$$

The obtained field coefficients (Eq. S17), when substituted into (Eq. S3) - (Eq. S6), allow us to visualize the fields (see Fig. 3d of the main text).

# Supplementary Note V. Dielectric function and growth parameters of pentacene

To obtain the dielectric function of pentacene, we measured reflectance spectra using Fourier-Transform Infrared Spectroscopy (FTIR) on a 150 nm layer of pentacene ($C_{22}H_{14}$) thermally evaporated on a $BaF_2$, and used the Tinkham formula (Eq. S18) for thin films on a substrate:

$$\frac{T}{T_0} = \frac{1}{\left|1+\sigma(\omega)d\frac{Z_0}{n+1}\right|^2} \quad (S18)$$

Here, $\sigma(\omega)$ is the complex conductivity of the thin film, which is related to the permittivity as $\varepsilon(\omega) = 1 + \frac{i}{\omega\varepsilon_0}\sigma(\omega)$. We have modeled the dielectric function using the Drude-Lorentz model with six oscillators to represent the molecular vibrations (Eq. S19):

$$\varepsilon_\infty + \sum_{i=1}^{6} \frac{s_i\omega_i^2}{\omega_i^2-\omega^2-i\gamma_i\omega} - 1 \quad (S19)$$

where $\varepsilon_\infty$ is the high-frequency dielectric constant (2.98), $s_i$ are the oscillator strengths, $\omega_i$ are the resonance frequencies, and $\gamma_i$ are the damping factors. The entire expression inside the absolute value is then squared and subtracted from 1 to relate it with the reflectance measurements obtained from the FTIR measurements (Eq. S20):

$$R = 1 - \left(\frac{1}{\left|1+\left(\varepsilon_\infty+\sum_{i=1}^{6}\frac{s_i\omega_i^2}{\omega_i^2-\omega^2-i\gamma_i\omega}-1\right)\left(\frac{\omega\epsilon_0 dZ_0 c}{i(n+1)}\right)\right|}\right)^2 \quad (S20)$$

We have also a transmission term (Eq. S21) involving the transmission characteristics:

$$\frac{\omega\epsilon_0 dZ_0 c}{i(n+1)} \quad (S21)$$

where $\epsilon_0$ is the permittivity of free space (8.85x10$^{-12}$ F/m), $d$ is the film thickness (150 nm), $Z_0$ is the impedance of free space (377 Ω), $c$ is the speed of light (3x10$^8$ m/s), and $n$ is the refractive index of the substrate (2.36).

The initial values for the fitting were obtained from previous ellipsometry measurements reported in the literature[5], allowing for accurate fitting. We have also subtracted the baseline of the experimental measurements, which was tilted due to scattering effects related to the roughness of the sample.

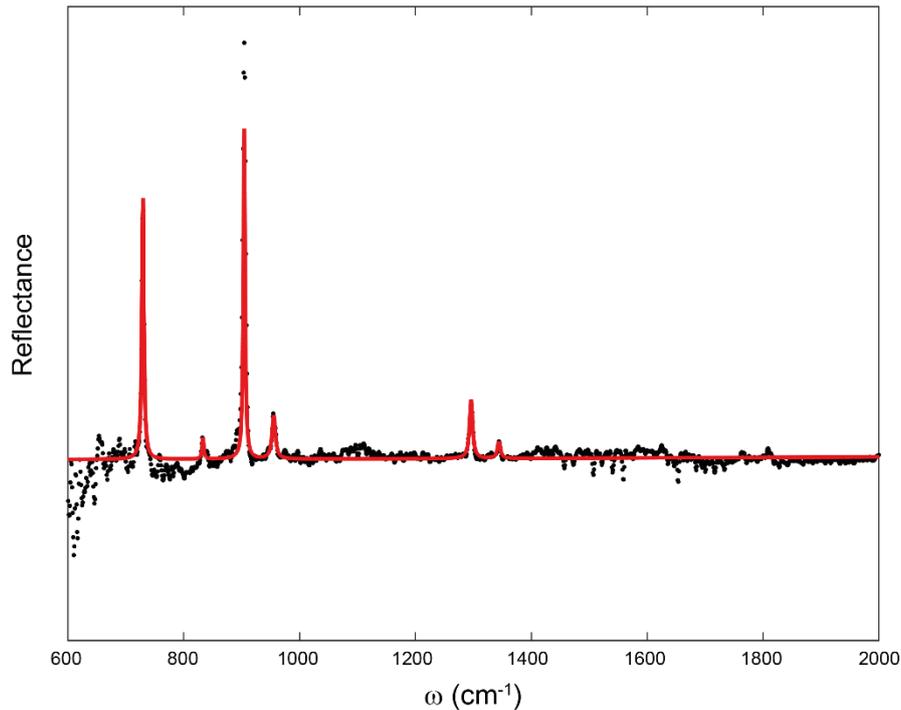

**Supplementary Fig. 12. FTIR reflectance spectra of pentacene molecules.** The black dots are the experimental points measured by FTIR, and the red solid line is the fitting from Eq. S20.

Supplementary Fig. 12 shows the reflectance spectrum of the 150 nm pentacene layer on $BaF_2$ and the fit using Eq. S20. All fit parameters are listed in Supplementary Table 1.

| $S_i$ | $\omega_0$ | $\Gamma_0$ |
|---|---|---|
| 0.0253 | 729.6 | 3.97 |
| 0.001189 | 833.2 | 3.127 |
| 0.018 | 904.4 | 3.5 |
| 0.00404 | 955.4 | 6.772 |
| 0.00268 | 1296 | 6.128 |
| 0.0006859 | 1344 | 5.937 |

**Supplementary Table 1. Parameters for the IR dielectric function of pentacene.** These parameters were extracted from the fitting model (Drude-Lorentz oscillator model) of the FTIR experimental data.

The dielectric functions of other materials used, such as $\alpha$-$MoO_3$ and $BaF_2$, have been taken from the literature[6].

The two pentacene samples, with thicknesses of 36 nm and 150 nm, used for the experiments were grown under identical conditions on $BaF_2$ substrates. The growth parameters were an evaporation temperature of 150 ºC, a deposition rate of 0.2 Å/s, and room-temperature growth. As shown in the AFM images and the extracted profiles in Supplementary Fig. 13, the films exhibit roughness, which could potentially affect the extraction of the polariton wavelength when mapping dispersion. This issue is particularly pronounced for smaller wavelengths, whereas it is less problematic for larger ones. It is worth noting that smoother films could improve measurement precision for smaller wavelengths, leading to more accurate extraction of dispersion points in the upper branch. One way to reduce roughness is by lowering the substrate temperature during growth. Growth at lower temperatures can enhance the molecular ordering of organic thin films during evaporation, resulting in a more uniform and smoother surface.

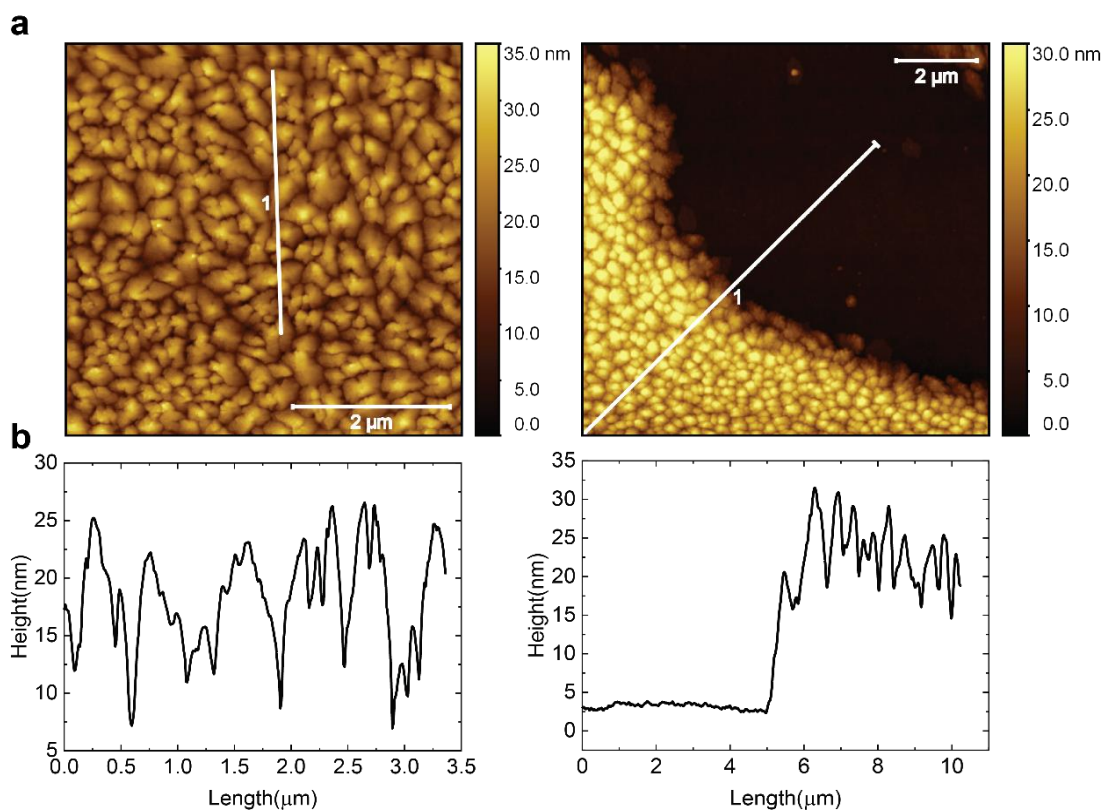

**Supplementary Fig 13. AFM characterisation of pentacene thin film. a**, AFM images of the 36 nm thick pentacene thin film. **b**, AFM profiles extracted along the white lines in **a**, illustrating the surface roughness of the thin film.

## Supplementary Note VI. Analytical study of the polaritonic fields

To deeply understand the strong coupling phenomena, we have analysed two different situations based on the pentacene layer thickness: a thick pentacene layer of 150 nm and a thin pentacene layer of 36 nm. We have plotted the analytical dispersion relations (see Eq. S11) for both cases in Supplementary Figs. 14 and 15, respectively, and compared them with a system having a semi-infinite layer of pentacene. A semi-infinite pentacene substrate show the maximum coupling that can be achieved between the PhPs and the molecules. Note that for the finite pentacene layer, part of the PhPs field might lie in the substrate, decreasing the maximum splitting.

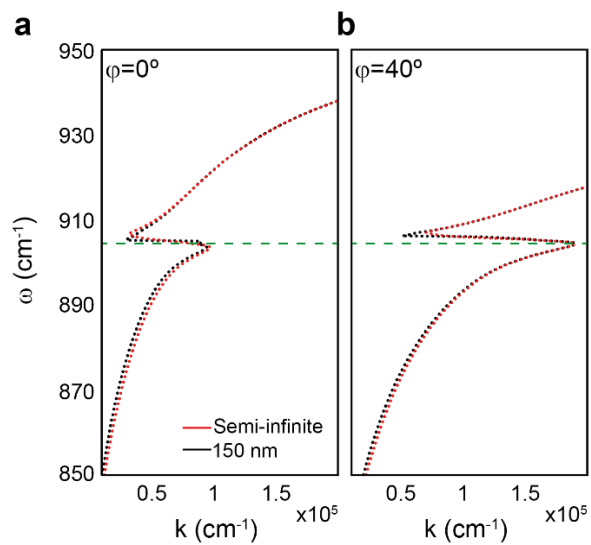

**Supplementary Fig. 14. Analytical dispersion relations for a thick pentacene layer.** The black dots represent system 1: $BaF_2$ - Pentacene (150 nm) - $\alpha$-$MoO_3$ (135 nm) - Air, while the red dots represent the semi-infinite system 2: Pentacene - $\alpha$-$MoO_3$ (135 nm) - Air. The analytical dispersions are shown for two different angles: $\varphi$=0 and 45 degrees, labeled as **a** and **b**, respectively.

In the case of a thick pentacene layer, as shown in Supplementary Fig. 14, we see that the dispersion of the system (black dots) is very similar to that of a system with a semi-infinite pentacene layer (red dots). This means that there is hardly any intensity below the 150 nm of the pentacene layer. In other words, the $BaF_2$ does not influence the system in the vicinity of the resonance frequency, where the absorption occurs. However, it should be noted that in the case of $\varphi$=0° (Supplementary Fig. 14a), some polaritonic field still reaches the substrate. Furthermore, the lower branch presents more differences than the upper one. This is consistent with the evolution of the penetration depth in Fig. 3 of the main text. The dispersion of the system with pentacene acting as a substrate (red dots) is outside the dispersion of the 150 nm layer (black dots). This means it shows higher (respectively lower) k values than those of the 150 nm layer below (respectively above) the pentacene resonance frequency. Therefore, the system with semi-infinite pentacene has a larger splitting and, consequently, so does its SC. Adding more molecules (considering a thicker pentacene layer) would slightly increase $\delta\omega$.

In the case of $\varphi$=40° (Supplementary Fig. 14b), the red and black lines coincide. This indicates that the 150 nm pentacene layer is effectively a semi-infinite medium in terms of the SC, and thus, the entire field is absorbed in the vicinity of the resonance frequency by the 150 nm layer of pentacene. Again, this aligns with the evolution of the penetration depth as a function of the angle in Fig. 3 of the main text.

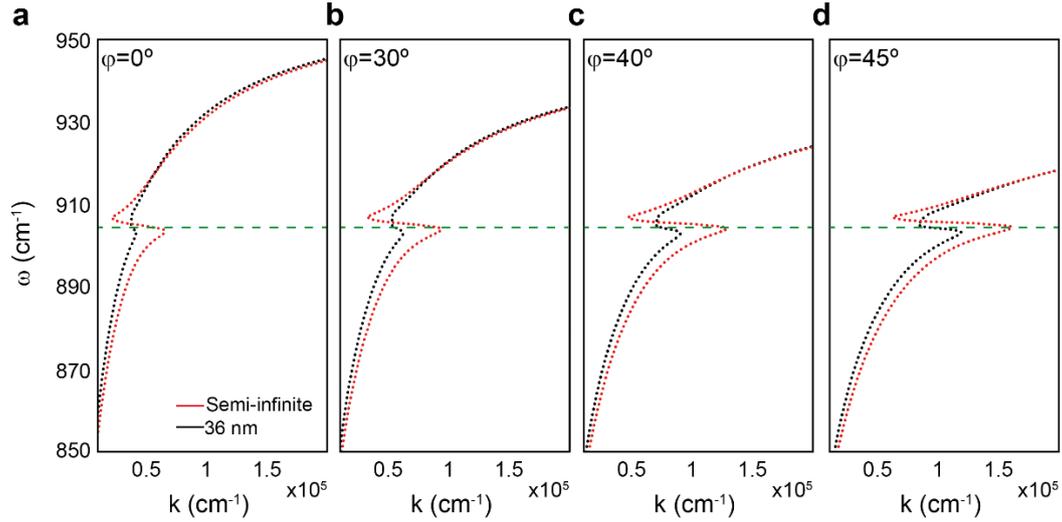

**Supplementary Fig. 15. Analytical dispersion relations for a thin pentacene layer.** The black dots represent system 1: $BaF_2$ - Pentacene (36 nm) - $\alpha$-$MoO_3$ (200 nm) - Air, while the red dots represent the semi-infinite system 2: Pentacene - $\alpha$-$MoO_3$ (200 nm) - Air. These analytical dispersions are shown for four different angles: 0, 30, 40, and 45 degrees, labeled as **a**, **b**, **c**, and **d**, respectively.

We now turn to the case of the thin film. We again compare the dispersion of this system with that of a semi-infinite pentacene substrate as a function of the in-plane angle. In this case, the pentacene layer is thin enough that much of the field extends beyond the pentacene layer, even at the resonance frequency. This unequivocally explains the significant difference observed between the red and black lines. It can be deduced that a large portion of the field is outside the slab, and therefore, we are far from achieving the maximum splitting (Supplementary Fig. 15). Thus, it is clear why there is a considerable difference between the splittings obtained at 0º for the thin (Supplementary Fig. 15a) and thick layers (Supplementary Fig. 14a) ($\delta\omega \simeq 8$ and $\delta\omega \simeq 15$, respectively). On the other hand, if we look at the evolution with the angle, we see that the difference between these two lines is reducing. Therefore, as the angle increases (Supplementary Fig. 15b), a greater portion of the field intensity available for that angle is increasingly contained in the pentacene layer. This causes the interaction with the molecules to be larger and, therefore, the splitting increases. Note that for the case of 45º (Supplementary Fig. 15d), there is still part of the field that comes out of the pentacene layer, although slightly (which fits with the penetration depth graph in Fig. 3c of the main text). This explains why the $\delta\omega$ values of the thin and thick cases do not coincide for 40º (Supplementary Figs. 15c and 14b). If the thin layer were a little thicker, both graphs should coincide from a sufficiently large angle (also assuming that the $\alpha$-$MoO_3$ layers had the same thickness).

## References

1. Dai, S. *et al.* Efficiency of launching highly confined polaritons by infrared light incident on a hyperbolic material. *Nano Lett.* **17**, 5285–5290 (2017).